\documentclass{azh2}

\usepackage{graphicx}
\usepackage{natbib}
\usepackage{lscape}

\def\degr{\hbox{$^\circ$}}
\def\arcmin{\hbox{$^\prime$}}
\def\arcsec{\hbox{$^{\prime\prime}$}}

\begin{document} 

\title{On the nature of spiral arms in the interacting galaxy M51}

\author{F.~Kh.~Sakhibov,$^1$
        V.~S.~Kostiuk,$^2$
        A.~S.~Gusev,$^2$
        and E.~V.~Shimanovskaya$^2$}

\institute{$^1$ University of Applied Sciences, Technische Hochschule  
                Mittelhessen, Friedberg, 
                Germany \\
           $^2$ Sternberg Astronomical Institute, Moscow State University, 
                Moscow, Russia}

\date{Received November 19, 2024; revised January 29, 2025; accepted Month 00, 2025}
\offprints{Firouz Sakhibov, \email{fsakhibov@yahoo.com}}

\titlerunning{On the nature of spiral arms in M51}
\authorrunning{Sakhibov et al.}

\abstract{In this study, we investigated the radial variation of the azimuthal propagation of star formation across the spiral arms in the nearby galaxy NGC 5194 (M51a) by analysing the spatial separation between young star clusters and their nearest H\,{\sc ii} regions. The significant differences in the radial profiles of the mean azimuthal offsets in the M51a arms were found when the southern and northern arms were studied separately. The northern arm analysis showed that its radial profile is consistent with the predictions of stationary density wave theory for trailing spirals, while the explanation of the radial profile for the southern arm required its pattern speed to be higher than the rotation velocity of disc matter within the corotation circle, and lower than that outside of it.
At the same time, these different radial profiles of the mean azimuthal offset in the two arms are joined by a common localization of corotation resonances, confirmed by independent studies using different methods. \\


{\bf DOI:} 10.1134/... \\
}

\maketitle

\section{Introduction}
\label{sect:intro}

Two alternative versions of the origin of the spiral structure of the interacting galaxy M51 were considered earlier: \\
1) the emergence of a gravitational quasi-stationary density wave in the galactic disc manifests itself as a stationary and long-lived spiral structure \citep{Lindblad1963, Lin1964, Bertin1989a, Bertin1989b}, \\
2) or as a transient feature caused by gravitational interaction with a close companion that suppresses any pre-existing structure \citep{Tully1974, Toomre1974}.

Most attempts to distinguish between these two theories have focused on the nearby grand-design spiral galaxy NGC 5194 (M51a), which is both a typical example of an interacting galaxy with its companion galaxy NGC 5195 (M51b) and, with a complex gravitational potential structure expressed in the complexity of its kinematics. Moreover, the large angular size and clearly visible, ordered structure of the spiral arms \citep{Elmegreen2011b}, with a large contrast in the brightness of the arms compared to the brightness of the inter-arm space, make the galaxy M51a ideal for studying their nature and testing theoretical models.
\citet{Tully1974}, after a kinematic study of {\sc M51}a, concluded that the spiral pattern in the outer disc, $R>5.7$~kpc ($>150\arcsec$), is a transient, material and non-wave structure caused by the interaction between M51a and its satellite M51b, while the inner segments of the spiral arms are generated by a quasi-stationary gravitational density wave, which is due to an external, material spiral.
Further studies of the radial amplitude profile of the spiral arms obtained in $B$ and $I$ bands \citep{Elmegreen1989}, the analysis of CO-radio maps and the determination of the spiral pattern rotation rate using the Tremaine-Weinberg method \citep[TW-method;][]{Zimmer2004, Meidt2008}, analytical study of a numerical experiment to investigate the excitation mechanisms of M51 spiral arms and the role of the halo for these mechanisms \citep{Howard1990}, 3D-modelling of the dynamical N-particle system \citep{Salo2000}, kinematics and dynamics of giant molecular clouds (GMCs) in the M51 disc \citep{Miyamoto2014, Colombo2013}, using high spatial resolution PAWS (PdBI Arcsecond Whirlpool Survey) velocity field observations, are consistent with the conclusion of \citet{Tully1974} about the dual nature of the spiral arms in {\sc M51}a: the wave nature of the inner segments of the spiral arms in {\sc M51}a and the non-wave outer spiral structure caused by gravitational interaction with the companion M51b. 
In case the conclusions of the above-mentioned investigations are valid, the radial modulation of star formation propagation by a quasi-steady spiral density wave should manifest itself only in the inner segments of the spiral structure, but not in the outer segments of the spiral arms.

The study of the azimuthal offset between objects of different ages (colour-age gradient) in spiral arms and its variation along the radius of the galactic disc is one of the widely used observational approaches for revealing the impact of the gravitational spiral density wave in star formation in nearby galaxies
\citep{Dixon1971, Smirnov1981, Efremov1985, Berdnikov1987, Beckman1990, Puerari1997, Gittins2004, Egusa2004, Zhang2007, Tamburro2008, Buta2009, Egusa2009, Martinez2009, Pour-Imani2016, Egusa2017, Shabani2018, Miller2019,Peterken2019,Sakhibov2021, Abdeen2020, Abdeen2022}.
A relatively comprehensive review of studies directly verifying density wave theory can be found in \citet{Peterken2019}.
The papers listed above consider colour-age gradients in nearby galaxies.
\citet{Martinez2023} recently discovered colour jumps and gradients in distant Hubble Ultra Deep Field galaxies. 

Below, we briefly recall the possible origin of the azimuthal age gradient in spiral arms and its dependence on galactocentric distance
A quasi-stationary gravitational density wave generates a galactic spiral shock wave in interstellar gas \citep{Fujimoto1968, Roberts1969, Pickelner1971}.
Since the rotational velocity of matter in the disc varies with galactocentric distance and the spiral pattern of the gravitational density wave rotates rigidly with a constant angular velocity at all galactocentric distances, the front of the galactic shock wave at some distance from the galactic centre changes its position relative to the spiral pattern:
it passes from the inner (concave) side of the spiral arm to the outer (convex) 
side if the spiral is trailing, or vice versa from the outer to the inner side if the spiral is leading.
In other words, if the spiral pattern of the galaxy does have a characteristic constant velocity, then we should expect that the angular azimuthal offset between any set of tracers, differing in age, will vary with galactocentric distance in a certain way.
Figure 1 in \citet{Puerari1997} clearly demonstrates the mutual arrangement of a two-arm spiral pattern and a stationary spiral shock in interstellar gas for two possible pattern directions (S-shaped or Z-shaped spirals), as well as for two possible rotation directions (clockwise or counterclockwise). 
The galactocentric distance at which the shock front changes position relative to the side of the spiral pattern and the rotation speed of the gas and stars in the disc coincides with the rotation speed of the spiral arms is called the radius of the corotation circle, $R_C$.
This, one of the basic theoretical predictions of the density wave theory about the mutual positioning of the shock front and spiral arms, was the main assumption for morphological methods of determining the corotation radius based on the geometrical measurement of the spatial angular displacement of different spiral structure tracers in galaxy discs in the studies mentioned above.

We note that the analysis of the radial profile of azimuthal offsets has revealed the multi-modal nature of the spiral structure in the galaxies NGC~628, NGC~3726 and NGC~6946 \citep{Sakhibov2021}.
The locations of the resonances in these galaxies match the corotational resonances for the first ($m=1$) and second ($m=2$) modes of the density wave derived using Fourier analysis of the observed line-of-sight velocity field \citep{Sakhibov2021}, as well as the estimates of the resonances obtained by the potential density phase shift method \citep{Buta2009} and by the Font-Beckman method using the change of sense of the radial component of the residual velocity at the resonance radius \citep{Font2011,Font2014}.
The overlap of the corotational radii, detected from an analysis of the radial profiles of azimuthally averaged offsets,  with the corotational resonances of the different modes of density waves, along with the strongly pronounced two-arm pattern as well as in a multiple spiral system, indicates the interference of these different modes of density waves and the multi-modal nature of the spiral structure found earlier in other galaxies \citep{Sakhibov1987, Elmegreen1989}.
The modal approach was further extended in \citet{Marchuk2024AA}, where cases of independent spiral patterns for galaxies with a flat rotation curve are identified and the relations between the relative rates of the patterns and the possible cases of coupling between the main resonances.

A similar approach, based on the study of the azimuthal colour-age gradient across the spiral arms, has been applied by several authors to study the spiral structure in the M51 galaxy considered in this paper \citep{Tamburro2008, Egusa2009, Louie2013, Egusa2017, Shabani2018, Abdeen2020}.
In the first of these papers, \citet{Tamburro2008} fitted, similar to the approach of \citet{Egusa2004}, the measured radial dependence of the angular offset between the spiral patterns seen in the distribution of neutral hydrogen HI and in the distribution of heated gas (24~$\mu$m) with a model assuming solid-state rotation of the spiral pattern with angular velocity $\Omega_p$ and a characteristic time span between the dense H\,{\sc i} phase and the formation of massive stars, which heat the surrounding dust and obtained that the radial dependence of the observed angular offset (H\,{\sc i} and 24~$\mu$m emission) is consistent with the model presumption as well as an estimate of the corotation radius at a distance of $R\approx4.77$~kpc ($\approx125\arcsec$). 

\citet{Egusa2009}, by the same method, studied the spatial angular offset between molecular and young-stellar arms M51a using CO and H${\alpha}$ images.
Since the offsets in the inner segment of the northern Arm~2, which is directly associated with the companion galaxy M51b (NGC~5195), show a negative dependence on the rotational angular velocity $\Omega$ of matter in the disc, the authors considered the offsets in the southern Arm~1 only and obtained an estimate of $R_C \approx6.65$~kpc ($\approx174\arcsec$).
Later, \citet{Egusa2017} measured the gas-star offsets from the location of gas density peaks relative to stellar density peaks in the two spiral arms, obtaining the opposite result of previous paper \citep{Egusa2009}, finding that only the northern Arm~2 associated with the companion galaxy NGC~5195 is consistent with a galactic shock wave for the estimate of the corotation radius $R_C\approx6.4$~kpc ($\approx168\arcsec$).

In this paper we have adopted the notation for the arms according to the notation in \citet{Shabani2018}, where the spiral arm that interacts with the companion galaxy NGC~5194 located in the north is called the ''northern Arm~2''. Consequently, the opposite arm is labeled as ''southern Arm~1''.
Some authors designate the arm interacting with the northern companion as the ''eastern arm'' or ''Primary Spiral Arm'', and the opposite arm as the ''western arm'' or ''Secondary Spiral Arm''.

\citet{Abdeen2020}, using the method of overlaying images of galaxies observed at different wavelengths and then tracing the spiral arms in them, determined the galactocentric distances where the spirals observed at different wavelengths intersect each other as the corotation radii of the spiral pattern and stars in the disc.
They localised the corotation radius in M51a at galactocentric distances between 4.3~kpc and 4.7~kpc ($113\arcsec$ and $123\arcsec$). 
In \citet{Scheepmaker2009}, the authors studied the azimuthal distribution of clusters in M51 in terms of kinematic age away from the spiral arms indicates that the majority of the clusters formed $\sim5-20$~Myr before their parental gas cloud reached the centre of the spiral arm.
\citet{Shabani2018} studied the age gradients of clusters across spiral arms in M51a using the stellar cluster catalogues from the Legacy Extragalactic UV Survey (LEGUS) program and found that this galaxy displays an offset in the location of young and old star clusters across southern Arm~1, while the authors did not observe an expected offset in the azimuthal distribution of star cluster samples across northern Arm~2. In contrast, in another paper \citet{Pineda2020} showed a clear offset between the CO and [C\,{\sc ii}] associated only with the northern arm, rather than with the southern one. Additionally, there are several papers that have found no significant angular offset between spiral features traced by stars of different ages \citep{Foyle2011, Kendall2011,Kendall2015, Marchuk2024}. It is worth noting studies~\citet{Pettitt2017} and~\citet{Dobbs2010}. The authors of the first paper found a clear offset between stars and gas, whereas the other did not find any sign of it, although both performed hydrodynamic simulations of a target galaxy including tidal interactions with its companion. Such disagreement in the presence of age gradients is not unique to M51 and is also observed for other spiral galaxies, as mentioned in~\citet{Vallee2020}. In addition, such discrepancies were also found in the estimation of corotation radii using different methods~\citep{Kostiuk2024}, which may indicate the existence of galaxies that follow both spiral formation scenarios. Besides, some difficulties in interpreting the obtained angular offsets may also arise due to the ambiguous mutual arrangement of stars of different ages within and outside the corotation circle. It should be noted that both of these pictures of age gradient propagation were found in real galaxies and are consistent with the density wave theory \citep{Miller2019,Martinez2023}.

In the current study, we use as tracers young star-formation complexes in spirals that emerged from giant molecular cloud complexes (GMCC) that experienced the impact of a galactic shock generated by a quasi-stationary spiral density wave.
These young star-formation complexes resulting from gas transformation in the parent GMCC contain spatially separated subgroups of extremely young OB-stars surrounded by giant H\,{\sc ii} regions and subgroups of older B-stars without ionised gas \citep{Gusev2019}. 
We explore the nature of spiral arms in the target galaxy M51 by measuring angular offsets between the photometric centres of the B-stars cluster and the region of ionised hydrogen in close ''star cluster--H{\sc ii} region'' pairs (SC--H{\sc ii}R) in young star formation complexes. 
We used a similar approach to study the azimuthal propagation of star formation in three nearby galaxies, NGC 628, NGC 3726 and NGC 6946 \citep{Sakhibov2021}.

There are several other mechanisms that may be responsible for the spatial offset between subgroups of stars of different ages. These may be stochastic self-propagating waves of star formation, which have no dedicated direction of propagation, but are possibly initiated by primary star formation caused by the effect of the galactic spiral shock or by processes caused by non-circular, radial motions of matter in the disc. Actually, contrary to the spiral structure, the spatial offsets of young star clusters with nearby H\,{\sc ii} regions measured in the galaxies NGC~628, NGC~3184, NGC~3726, NGC~5585, and NGC~6946 show trends suggesting that spiral density waves are not the dominant mechanism of star formation in them \citep{Gusev2019}. On the other hand, the global spiral patterns in Grand Design-type galaxies, to which the NGC 628 galaxy studied in \citet{Gusev2019} belongs, are undoubtedly caused by a spiral density wave. Although the contribution of the spiral shock wave to the spatial offset in SC--H{\sc ii}R pairs is hidden in the galaxies studied in \citet{Gusev2019} by the action of other mechanisms initiating star formation, it can be and has been revealed in \citet{Sakhibov2021} through the characteristic radial profile of the mean magnitude of the azimuthal component of the spatial offset of young star clusters from nearby H\,{\sc ii} regions. Additionally, we note that the contribution of the spiral shock wave to star formation was also recognized in \citet{Marchuk2024b} by constructing a photometric azimuthal profile in spiral arms not only in nearby but also in distant galaxies up to $z\sim 0.9$.

The paper is structured as follows. 
In Section~\ref{sect:method} we describe the method of the morphological analysis of spatial separations between a star cluster and the nearest H\,{\sc ii} region 
(SC--H{\sc ii}R pair) in the studied galaxy. 
The data are discussed in Section~\ref{sect:data}. 
Results are presented in Section~\ref{sect:result}. 
Section~\ref{sect:discussion} discusses the results and outstanding issues.
Section~\ref{sect:conclusion} summarizes our conclusions.

\section{The Method}
\label{sect:method}

Let us present a qualitative picture of star formation in giant molecular clouds complex (GMCC), consisting of one or more molecular clouds and 
 which catches up with and passes through a spiral shock.
When such a GMCC in the disc overtakes the spiral arm, a spiral shock wave compresses it at the front edge and can induce star formation in the region of compression.
If the GMCC is quite large ($\ge100$~pc) and the velocity of the complex passing through the shock front is equal to the disc and spiral tangential velocity difference, it may take the GMCC several million years to pass through the galactic spiral shock.
This starts a process of new star formation in the region of compression at the front of the GMCC and, as the complex passes through a spiral shock, the compression process, and with it the formation of new stars, propagates along the GMCC.
A similar mechanism for the formation of successive generations of star clusters in a molecular complex was first proposed in \citet{Elmegreen1977}, where the authors propose that subgroups of OB stars are formed during a stepwise process involving the propagation of ionisation (I) and shock (S) fronts through a complex of molecular clouds. OB stars formed at the edge of the molecular cloud drive these I-S fronts inside the cloud. This sequential mechanism may account for the spatial separation and systematic differences in the age of OB subgroups in a star-forming region.
Thus, the stars formed at the initial moment when the front edge of the GMCC passed through the spiral shock, and the stars formed after the entire complex passed through the spiral shock, have an age difference proportional to the GMCC passage time through the spiral shock, and the direction of star formation propagation is parallel to the direction of disc rotation.
In this way, it could create an azimuthal age gradient for stars within a star-forming complex emerging from a single GMCC.

\begin{figure}
\vspace{1.0mm}
\centerline{\includegraphics[width=0.48\textwidth]{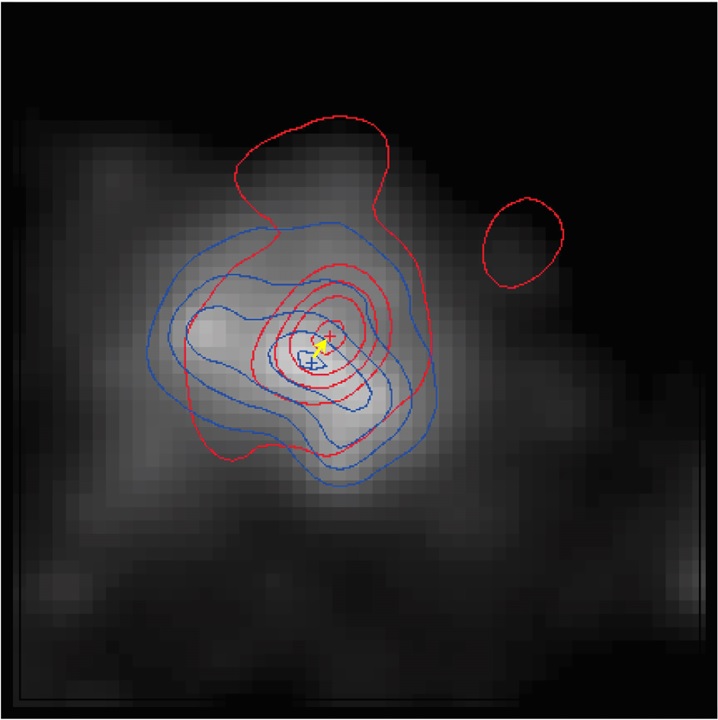}}
\caption{$B$ image of the star cluster -- H{\sc ii} region pair in M51 with overlaid isophotes in the H$\alpha$ line (red) and $B$ band (blue). The photometric centres of the star cluster (blue cross) and the H\,{\sc ii} region (red cross) are connected by a yellow arrow indicating the direction of star formation. The image covers a sky area of $15.8\times15.8$~arcsec$^2$, which corresponds to a linear scale of $600\times600$~pc$^2$.}
\label{figure:example}
\end{figure}

Newly born stars are usually obscured by dust and H\,{\sc ii} regions formed by short-lived massive OB stars. For about the first 10 Myr, supernova outbursts and the disappearance of the most massive and hot stars that can ionise interstellar hydrogen remain a cluster of less massive B stars without ionised gas in the location of the OB-star cluster. 
With a difference in matter and spiral shock tangential velocity of 5 to 100 km/s and an age difference of 10 million years between newborn OB stars and a subgroup of less massive B stars without ionised gas, the GMCC moves from 50 to 1000 pc and the star formation complex itself, consisting of successive generations of star clusters emerging from one giant molecular gas-dust complex, can leave the region of the spiral arm. 
Figure~\ref{figure:example} shows an example of spatial offset between a subgroup of extremely young OB stars surrounded by ionised hydrogen and a subgroup of relatively older B stars without ionised hydrogen in the star-formation complex in the galaxy M51a.

\begin{table*}
\begin{center}
\caption{Global parameters of the M51 galaxy.}
\label{table:sample}
\begin{tabular}{c|c|c|c|c|c|c}
\hline\hline
Galaxy     & Type       & D$^1$ & Inclination$^2$ & PA$^3$ & $R_{25}$$^4$    &Number  \\
          &           & [Mpc] &  [degree]    &  [degree]  &  [arcmin] & of pairs   \\
1         &  2        &  3   &   4        &     5     &     6   & 7  \\
\hline
NGC~5194   &  SA(s)bc & 7.88  &  24       &    170     &   5.61   &  612 \\
\hline
\end{tabular}
\end{center}
\begin{flushleft}
Columns 1, 2: Galaxy name and morphological type as listed in the NASA Extragalactic Database (NED)\\
$^1$NED average distance estimation to M51 \\ 
$^2$ Inclination of the M51 according to \citet{Oikawa2014} \\
$^3$Position angle of the major axis according to \citet{Tully1974}\\
$^4$Radius of M51 at the isophotal level 25 mag\,arcsec$^{-2}$ in the $B$ band \citep{Elmegreen2011b}
\end{flushleft}
\end{table*}

Spiral shock-induced star formation can be a trigger for a stochastic self-propagating star formation (SSPSF) wave, caused by additional cloud compression in supernova explosions, or a supersonic expansion of H\,{\sc ii} zones (ionisation fronts) around hot OB stars and under the action of stellar winds \citep{Mueller1976, Seiden1982, Elmegreen2011}. In contrast to star formation induced by a large-scale spiral shock wave when a GMCC passes through a shock front and propagates in the tangential direction, stochastic star formation can propagate in any direction. 
Therefore, despite the dominance of stochastic star formation in the galaxy, the contribution of the galactic shock wave to star formation can be revealed by identifying and analysing the azimuthal component of the spatial offset between a subgroup of extremely young OB-stars surrounded by H\,{\sc ii} regions and a subgroup of older B stars without ionised hydrogen \citep{Smirnov1981, Sakhibov2021}.
Since azimuthal offsets between young star clusters and nearby H\,{\sc ii} regions within the same star-forming region trace star formation initiated by a gravitational spiral density wave in the disc, the presence or absence of regular traces of such azimuthal offsets may clarify the controversial picture of the origin of the grand design in the spiral, interacting galaxy M51.
In our previous papers \citep{Gusev2019,Sakhibov2021}, we have shown that although SSPSF processes dominate in the spiral galaxies NGC 628, NGC 3726, and NGC 6946 studied there, the impact of a stationary spiral density wave is nevertheless evident in the characteristic radial profile of the azimuthal component of the age gradient in young star-formation complexes or the azimuthal component of the spatial separation of young star clusters from nearby H\,{\sc ii} regions.

Under the dominating SSPSF process, to reveal the contribution of the spiral density wave to star formation, we considered only the azimuthal (tangential) component of the spatial offset between the photometric centres of star-forming regions in galaxy images obtained with the broadband $B$ filter and the narrowband H$\alpha$+[N{\sc ii}] interference filter in close SC--H{\sc ii}R pairs, using Eq. (1) from \citet{Sakhibov2021}:
\begin{equation}
\Delta_{azimuth} = (\varphi_{SC} -\varphi_{\rm HII})\cdot R_{SC},
\label{equation:azimuthal_offset}
\end{equation}
where $\varphi$ and $R$ are the polar coordinates of the objects in the deprojected coordinate system. 
We chose a right-hand polar coordinate system in which the polar angle $\varphi$ is counted in the counterclockwise direction and the polar axis is directed from the centre of the deprojected disc along the northern part of the major axis of the galaxy.
South of the centre, along the major axis, the polar angle is $\varphi = 0\degr$, and west of the centre, along the minor axis, is $\varphi = 90\degr$. 
Lower indices indicate the type of object: ''$SC$'' for star cluster coordinates, ''H\,{\sc ii}'' for H\,{\sc ii} region coordinates.
We define the azimuthal offset using Eq.~(\ref{equation:azimuthal_offset}) so that its sign will be positive for S-shaped trailing spirals and negative for Z-shaped trailing spirals inside the corotation resonance, with the gas shock located on the concave side of the spiral.
Outside the corotation resonance, the signs of the azimuthal offset will be reversed, and the gas shock will be located on the convex side of the spiral.

As noted above, the impact of the quasi-stationary density wave on star formation is manifested in the radial variation of the magnitude and direction of the azimuthal offset between tracers of different ages.
In order to reveal the radial modulation of the azimuthal offset between subgroups of stars of different ages born in the molecular complex, the galaxy disc was divided into thin annuli with width $\Delta R=3$~kpc. For each annulus, a mean value of the azimuthal offset was calculated and its behavior with respect to changes in galactocentric distance was analysed.
To compensate for the effect of non-systematic random errors in the case of a small number of pairs within the individual annulus, we applied smoothing using the weighted moving average (WMA) method.
Meanwhile, the galaxy radius step from one annulus to the next one was chosen to be three to ten times smaller than the annulus thickness, so that neighbouring annuli overlap partially.

If there is a radial modulation of the direction of star formation propagation by the spiral shock, the sign of the mean azimuthal offset calculated from Eq.~(\ref{equation:azimuthal_offset}) is reversed when passing through the corotation radius of matter and the spiral density wave pattern in the disc.
In the case of an S-shaped trailing spiral (the galaxy rotates counterclockwise), the sign of the azimuthal offset calculated according to Eq.~(\ref{equation:azimuthal_offset}) will be positive inside the corotation circle and negative outside. 
In the case of an S-shaped leading spiral, if the galaxy rotates in a clockwise direction, the behaviour of the azimuthal offset sign will be opposite.

\section{The Data}
\label{sect:data}

The global parameters adopted in this paper for the target galaxy M51 are listed in Table~\ref{table:sample}.
To reveal the systematic variation of the azimuthal offset between the above-mentioned subgroups of stars of different ages in star-formation complexes (hereafter denoted as SC--H{\sc ii}R pairs) as a function of galactocentric distance, we measured the coordinates of the photometric centres of star clusters and ionised hydrogen regions from $B$ and H$\alpha$ images of M51, obtained on the 2.1-meter KPNO telescope at Kitt Peak National Observatory (Arizona, USA) by \citet{Kennicutt2003}. The images were downloaded via NED database.\footnote{http://ned.ipac.caltech.edu/}.

\begin{table*}
\begin{center}
\caption{Summary of estimates of corotation radii in M51a in comparison with results obtained by other authors.}
\label{table:summary}
 \renewcommand*{\arraystretch}{1.2}
 \begin{tabular}{c|c}
\hline\hline
  Present study  & Other studies  \\
\hline
$R_{C_1, Arm1}=0.90\pm0.15$ kpc $(24\pm4\arcsec)$ & $R_{C, bar} = 0.84\pm0.10^a$ kpc  $(22\pm2.5\arcsec)$   \\
         & $R_{C_1} = 0.81\pm0.12^b$ kpc     $(21.3\pm3.1\arcsec)$  \\
                             & $R_{C, m=3} \approx 1.15^c$ kpc  ($\approx30\arcsec$)   \\
                             & $R_{C, m=3} \le 1.6^d$ kpc    ($42\arcsec$)    \\
                             & $R_C = 1.19\pm0.14^e$ kpc       $(31.2\pm3.6\arcsec)$   \\
\hline
$R_{C_2, Arm1} = 3.70\pm0.15$ kpc $(97\pm4\arcsec)$ & $R_{C, spiral} = 3.9\pm0.4^a$ kpc   ($102\pm10\arcsec$)  \\
        & $R_{C_2} = 4.206\pm0.004^b$ kpc           $(110.1\pm0.1\arcsec)$   \\ 
 \hline                    
   $R_{C_1, Arm2} = 0.90\pm0.15$ kpc $(24\pm4\arcsec)$ & $R_{C, bar} = 0.84\pm0.10^a$ kpc  $(22\pm2.5\arcsec)$   \\
           & $R_{C_1} = 0.81\pm0.12^b$ kpc    $(21.3\pm3.1\arcsec)$  \\
                                & $R_{C, m=3} \approx 1.15^c$ kpc    ($30\arcsec$)   \\
                                & $R_{C, m=3} \le 1.6^d$ kpc    ($42\arcsec$)    \\
                                & $R_C = 1.19\pm0.14^e$ kpc       $(31.2\pm3.6\arcsec)$   \\
\hline
$R_{C_2, Arm2} = 4.50\pm0.15$ kpc $(118\pm4\arcsec)$   &   $R_C \approx 4.7^f$ kpc               ($123\arcsec$)      \\ 
           &  $R_C \approx 4.2^g$ kpc          ($110\arcsec$)      \\ 
                               & $R_C \approx 4.8^h$  kpc            ($126\arcsec$)        \\                         
                                & $R_C = 4.7\pm0.9^i$ kpc              $(123\pm23\arcsec)$   \\
                               & $R_C = 4.3\pm1.0^j$ kpc             $(113\pm26\arcsec)$   \\
                               & $R_C = 4.78\pm1.97^k$ kpc           $(125\pm52\arcsec)$   \\ 
                               & $R_C \approx4.78^l$ kpc 
                              ($125\arcsec$)   \\
                               & $R_C = 4.96\pm0.16^e$ kpc           $(130.0\pm4.1\arcsec)$   \\          & $R_{C_2} = 4.206\pm0.004^b$ kpc           $(110.1\pm0.1\arcsec)$   \\
\hline
\end{tabular}
\end{center}
\begin{flushleft}
\footnotesize{The values of the corotation radius $R_C$ found by other authors are rescaled according to the $R_{25}=5.61\arcmin$ and the distance $d=7.88$~kpc assumed in the current paper.\\ 
$^a$ The corotation radius (CR) was determined in \citet{Querejeta2016} from the radial profiles of azimuthally averaged gravitational torques: the torque profile shows a sign change at $R=22\arcsec$, which can be associated with the CR of the nuclear bar and at $R\approx100\arcsec$, which is suggested to indicate of the corotation of the spiral pattern. \\
$^b$ Corotation radii from potential-density phase shifts \citep{Zhang2012}. \\
$^c$ The corotation radius of the $m=3$ mode detected in \citet{Colombo2013} from molecular gas kinematics.\\
$^d$ From the pattern speed estimate calculated by the radial TW (TWR) method using CO(1-0) as a kinematic tracer \citep{Meidt2008}. \\
$^e$ The corotation radius is determined from reversals in the streaming motion caused by the spiral density wave \citep{Font2024}. \\
$^f$ Fitting the spiral pattern of M51a using density wave theory \citep{Tully1974}.\\
$^g$ \citet{Zimmer2004} applied the TW method to the CO map of the M51, obtaining a pattern speed of $38\pm7$ km/s/kpc, which leads to the corotation radius mentioned above. \\
$^h$ The corotation radius is determined from reversals in the streaming motion caused by the spiral density wave \citep{Vogel1993}. \\
$^i$ The corotational radius is determined by overlaying images of the galaxy observed at different wavelengths and identifying the locations where the traced arms intercept each other \citep{Abdeen2020}. The tracing of spiral arms was done using Python OL Script (https://github.com/ebmonson/2DFFTUtils-Module). \\
$^j$ The corotational radius is determined by the same overlaying technique \citep{Abdeen2020}, but the tracing of spiral arms was done using Spirality’s SpiralArmCount script \citep{Shields2015}.\\
$^k$ Corotation radius $R_C$ determined from the azimuthal offsets between the H{\sc i} and 24~$\mu$m emission peaks in the spiral arms \citep{Tamburro2008}. \\
$^l$ Corotation radius $R_C$ determined from azimuthal offsets between the peaks of stellar mass and gas mass distributions in spiral arms \citep{Egusa2017}.}
\end{flushleft}
\end{table*}

At the first stage, we identified and selected stellar clusters and star-forming regions in the $B$ and H$\alpha$ images of the target galaxy using the {\sc SExtractor}\footnote{http://sextractor.sourceforge.net/} program. The procedure of identification and selection has been described in detail previously in \citet{Gusev2018}. To ensure an approximately equal number of stellar clusters and star-forming regions during the selection process, we varied the magnitude threshold for both samples. Next, we determined the nearest H\,{\sc ii} region for each selected cluster using the deprojected coordinates of the samples. Note that we excluded pairs with distances less than the pixel size (0.3 arcsec). Note that, despite the availability of catalogs of H{\sc ii} regions \citep[see, for example,][]{Gutierrez2011,Honig2015}, we used our own list of these regions. This allows us to investigate SC--H{\sc ii}R pairs, the positions of which were determined using the same method.

\begin{figure*}
\vspace{1.0mm}
\centerline{\includegraphics[width=1.0\textwidth]{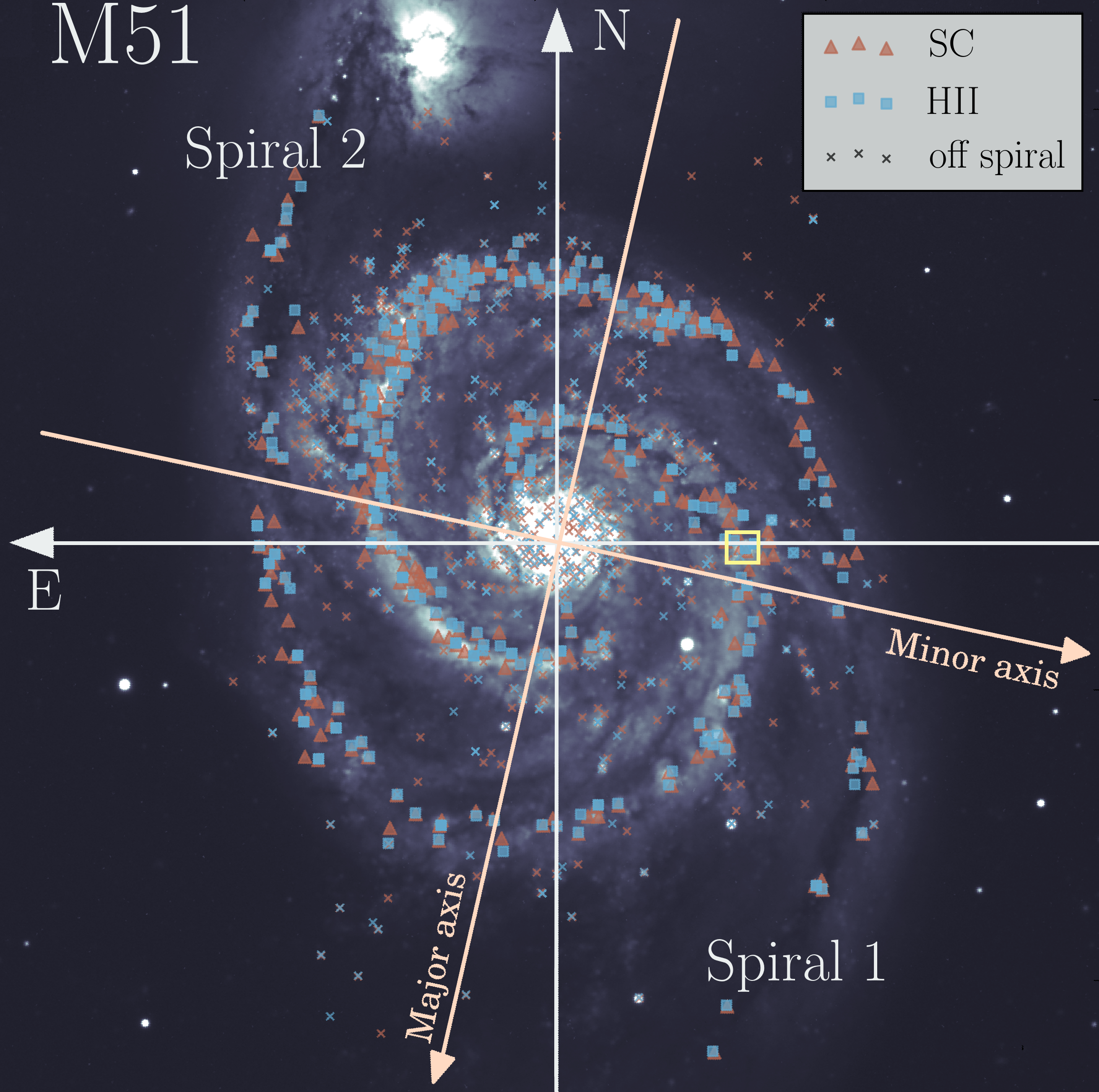}}
\caption{Map of selected star clusters (red) and H\,{\sc ii} regions (blue) in spiral arms of M51. Triangles and squares correspond to SC and H\,{\sc ii} regions located in spiral arms, while crosses indicate clusters occurring in off-spiral regions of the galaxy. The yellow square marks the region containing the star cluster -- H{\sc ii} region pair presented in Fig.~\ref{figure:example}.}
\label{figure:Map_M51}
\end{figure*}

In order to determine whether a particular SC--H{\sc ii}R pair corresponds to spiral arms, we used a mask of the spiral structure based on the $B$-band image. The mask was created by analysing slices across the spiral arms at different points in the spiral structure \citep[for a detailed description of this process, see][]{Savchenko2020}. Therefore, every stellar clusters (and their corresponding H{\sc ii} regions) located within the mask of the southern or northern arms were considered to be part of spiral 1 or spiral 2, respectively, while the other pairs were classified as ''off-spirals'' clusters.
Figure~\ref{figure:Map_M51} shows an image of the M51 galaxy in the $B$ band, with a non-deprojected map of selected star clusters (red markers) and H{\sc ii} regions (blue markers). The SC and H\,{\sc ii} regions located in the spiral arms are represented by triangles and squares, respectively, while the off-spiral regions are indicated by crosses.

\begin{figure*}
\vspace{1.0mm}
\centerline{\includegraphics[width=1.0\textwidth]{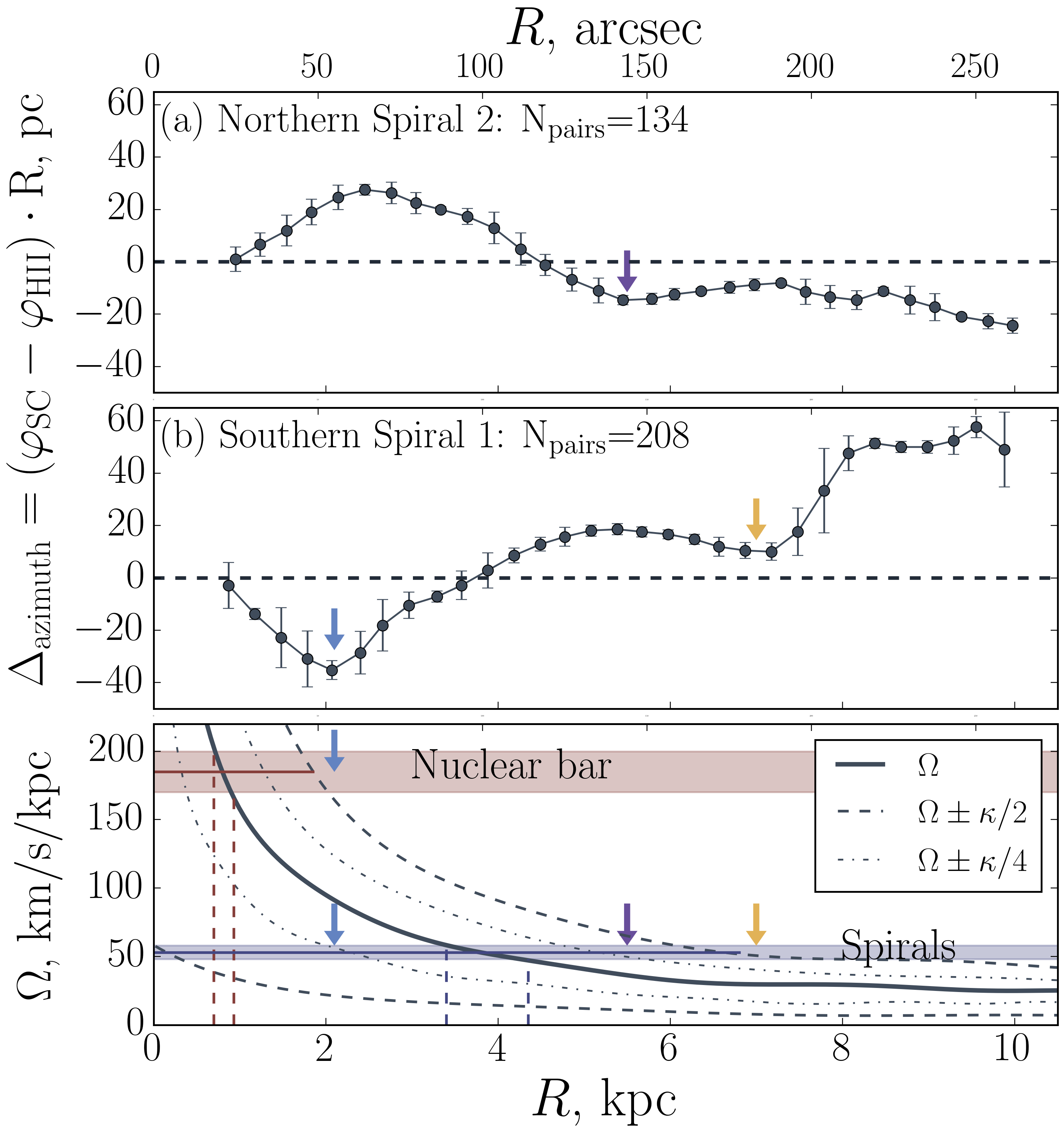}}
\caption{The top and central panels demonstrate the radial variation of the ''weighted moving average'' of the azimuthal offset in the SC--H{\sc ii}R pairs in the northern (a) and southern (b) arms in M51a, calculated in annular zones with a width of 3 kpc and a step of 0.3 kpc. Weighted average errors were specified as errors. The bottom panel displays the profiles of the disk angular velocity and the 1$\div$2 and 1$\div$4 resonances, taken from \citet{Querejeta2016}. The red and blue horizontal lines correspond to the pattern speeds of the nuclear bar and spirals, respectively.}
\label{figure:offset_M51}
\end{figure*}

\section{Results}
\label{sect:result}

As described in Section~\ref{sect:method}, to investigate the relationship between the azimuthal offset in the SC--H\,{\sc ii} pairs and the galactocentric radius, the deprojected galaxy disc was divided into concentric annular zones with a thickness of 3 kpc and a spacing of 0.3 kpc.
For every annular zone, we calculated the mean azimuthal offset and examined its behaviour as the galactocentric distance varies. 
This procedure was carried out for two different samples of SC--H\,{\sc ii} pairs in the galaxy disc.
The first sample included 134 pairs from the northern spiral of the galaxy M51a, hereafter designated as Spiral~2; the second sample included 208 objects located in the southern spiral, hereafter designated as Spiral~1, according to the identification of spirals in the studies of other authors \citep{Egusa2009, Egusa2017, Shabani2018}.
Recall that spiral Arm~1 leads south on the sky plane, and spiral Arm~2 leads north and towards the companion galaxy M51b.

In Fig.~\ref{figure:offset_M51}, we inspected the variation of the averaged azimuthal offset in SC--H\,{\sc ii} pairs, calculated using the Eq.~(\ref{equation:azimuthal_offset}), in concentric annuli as a function of the mean radius of the annulus in the deprojected disc of M51. 
As noted in Section~\ref{sect:method}, we chose a right-hand polar coordinate system, in which the polar axis is directed from the centre of the deprojected disc along the southern part of the major axis of the galaxy, with position angle PA$=170\degr$ and inclination $i=24\degr$, according to \citet{Oikawa2014}.
Meanwhile, the rotation angle of the celestial coordinate system, with the X-axis pointing West and the Y-axis pointing North in the plane of the sky, was $90\degr+$PA, i.e. $260\degr$.
In this way, in the deprojected coordinate system the X-axis coincides with the major axis of the galaxy with position angle PA $=170\degr$ (see Fig.~\ref{figure:Map_M51}).

The upper panel (a) of Fig.~\ref{figure:offset_M51} shows the radial variation of the ''weighted moving average'' (WMA) of the azimuthal offset in the concentric annular zones calculated for 134 SC--H{\sc ii} pairs in the northern Spiral~2, directly associated with the companion galaxy {\sc M51}b; the bottom panel (b) shows the radial variation of the WMA of the same quantity calculated for 208 SC--H{\sc ii} pairs in the southern Spiral~1.

By comparing panels (a) and (b) in Fig.~\ref{figure:offset_M51}, we can see that there are opposite trends in the radial profile of the mean azimuthal offset in the northern and southern spirals of M51a.
Earlier \citet{Egusa2017}, using their measurements of the gas-star offsets from the location of gas density peaks relative to stellar density peaks, found opposite radial dependencies for Arm 1 and Arm 2 at $R=1.15-5.73$~kpc $(30-150\arcsec)$.
Reminder that \citet{Shabani2018} also found an opposite age gradient in the northern Arm~2.
We discuss the significance of this result as well as the specific features of the radial profiles of the azimuthal offset magnitude in different arms in Section~\ref{sect:discussion}.

But consider, first, the radial profiles of azimuthal offsets in each arm separately.
The upper panel (a) in Fig.~\ref{figure:offset_M51} shows that the azimuthal offset measured in northern Spiral 2 varies with increasing galactocentric distance as follows: \\
-- in the central region of the disc, $R<1$~kpc ($<26\arcsec$), the mean azimuthal offset is not evident; \\
-- between galactocentric distances of 1 kpc and 5.4 kpc ($26\arcsec$ and $141\arcsec$), the positive azimuthal offset increases from zero to a positive peak at $R=2.43$~kpc, then it drops to the galactocentric distance of $\approx5.4$~kpc ($141\arcsec$), changing sign from positive to negative at 4.5~kpc ($118\arcsec$); \\
-- there is a kink in the radial profile of the negative amplitude of the azimuthal offset at $R\approx5.4$ kpc, the negative growth of the amplitude of the mean azimuthal offset is replaced by its quasi-constant value up to the distance $R=8.7$ kpc ($228\arcsec$), after which further growth of the negative amplitude is observed up to the investigated boundary of the disc at the distance $R=10$~kpc.

Consider further the radial variation of the azimuthal offset in the southern spiral Arm~1. The panel (b) in Fig.~\ref{figure:offset_M51} shows: \\
-- in the inner segment of the Arm~1, between galactocentric distances of 1 kpc and 3.7 kpc ($26\arcsec$ and $97\arcsec$), the negative azimuthal offset decreases from zero to a negative peak at $R=2$~kpc ($63\arcsec$), then it grows up to 4 kpc, changing sign from negative to positive at 3.7 kpc ($97\arcsec$); \\
-- at galactocentric distances $R>4$ kpc, the positive growth of the amplitude of the mean azimuthal offset is replaced by its quasi-constant value up to the distance $R=7$~kpc ($183\arcsec$); \\
-- there is a kink in the radial profile of the positive amplitude of the azimuthal offset at $R=7$~kpc, after which the amplitude continues to grow up to the investigated boundary of the disc at $R=10$~kpc ($260\arcsec$).

Summarising, it should be noted that the radial profiles of the mean azimuthal offset in the M51 arms have opposite trends, as well as opposite signs.
A similar result obtained in the study of \citet{Egusa2017} which, by measuring the azimuthal offsets between the peaks of the gas mass and stellar mass distributions, also found opposite signs and opposite trends in the two spiral arms. This is illustrated in table 3 and fig.~7 of \citet{Egusa2017}.
Previously, \citet{Egusa2009} using measurements of the azimuthal offset between the molecular (CO) and young stellar arms in the inner part of the galaxy in the interval $R=1.3-3.8$~kpc separately for the northern and southern spirals, found opposite dependencies of the magnitude of the azimuthal offset on the rotational velocity of matter in the disc in these arms.

\begin{figure*}
\vspace{1.0mm}
\centerline{\includegraphics[width=1.0\textwidth]{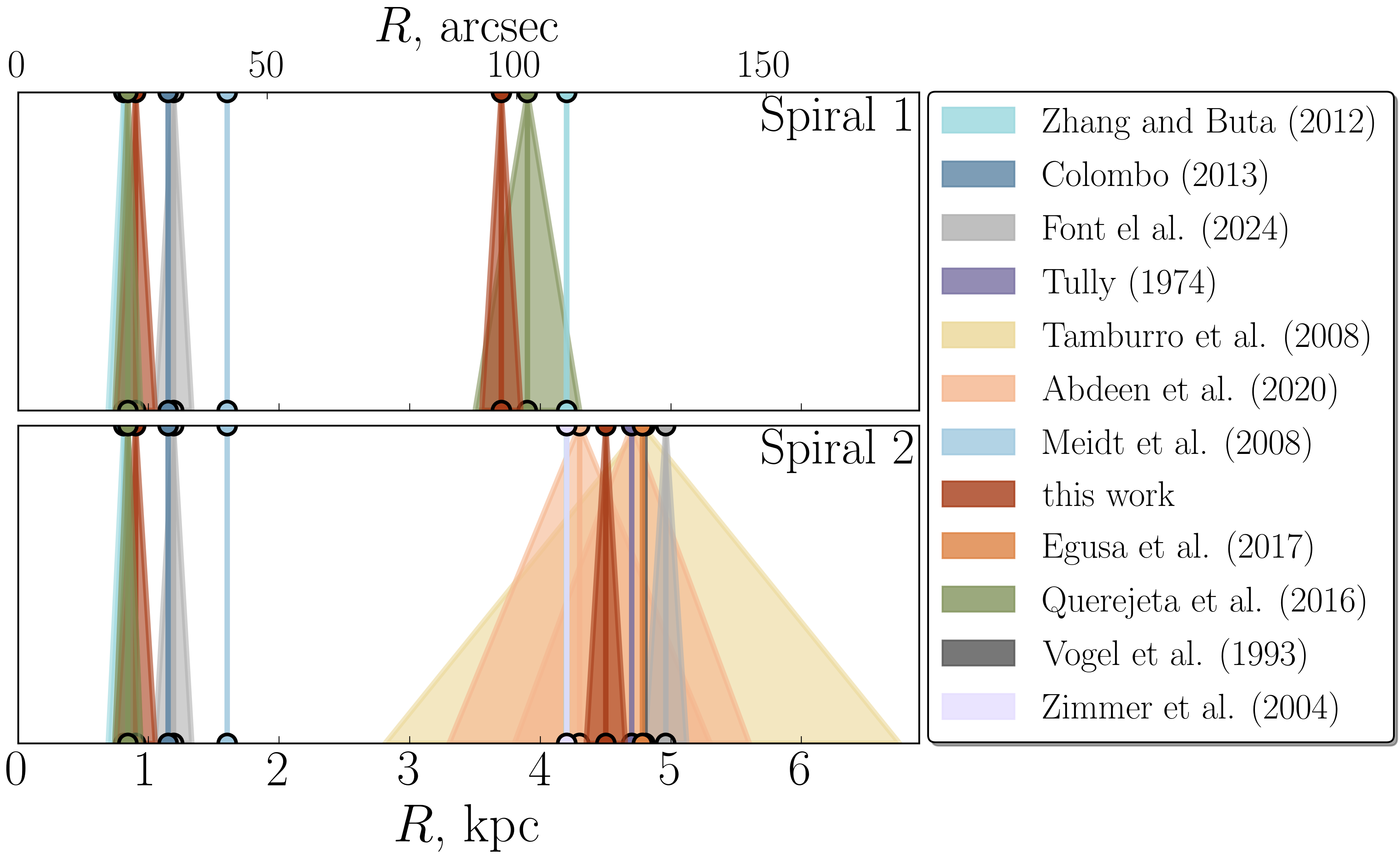}}
\caption{The distribution of corotation radii, as measured in this study and other studies, is shown in the top panel for the southern spiral arm and in the bottom panel for the northern spiral arm. The shaded areas represent the uncertainty in the measurements (the base of the cones). The color of each cone indicates the paper in which the particular value was obtained (see the legend on the right). The lower scale indicates their positions in kiloparsecs, while the upper scale is in arcseconds. A detailed description of each measurement is presented in Table~\ref{table:summary}.}
\label{figure:consistency}
\end{figure*}

\section{Discussion}  
\label{sect:discussion}

As promised in Section~\ref{sect:result}, we return to a more detailed discussion of the opposite trends in the radial profile of the averaged azimuthal offset in pairs in the northern and southern spiral arms and discuss their nature.
We first consider the radial profile of the azimuthal offset in northern Spiral~2 (Fig.~\ref{figure:offset_M51}a).
The radial profile of azimuthal offset in the northern Arm~2 is traced, starting from the region occupied by the end of the nuclear bar \citep{Querejeta2016,Colombo2014} at $R\approx0.8$ kpc ($\approx20\arcsec$). 
The zero value of the averaged azimuthal offset may indicate the existence of a corotation resonance, $R_C=0.90\pm0.15$ kpc ($24\pm4\arcsec$), associated with this central structure.
This estimate of the inner resonance is consistent with the corotation radius (CR) determined in \citet{Querejeta2016} from the radial profiles of azimuthally averaged gravitational torques: the torque profile shows a sign change at $R=22\arcsec$, which can be associated with the CR of the nuclear bar (see red line at the bottom panel of Fig.~\ref{figure:offset_M51}), 
with the CR determined from potential-density phase shifts \citep{Zhang2012},
with the CR of the $m=3$ mode detected in \citet{Colombo2013} from molecular gas kinematics,
with the estimate of the pattern speed derived in \citet{Meidt2008} using the radial Tremaine-Weinberg method (TWR-method) using CO(1-0) as a kinematic tracer,
and with the corotation radius determined in \citet{Font2024} from reversals in the streaming motion caused by the spiral density wave (see Table~\ref{table:summary} and Figure~\ref{figure:consistency}).

The further growth of the mean azimuthal offset up to the positive peak at $R\approx2.5$ kpc ($\approx65\arcsec$) and its decrease beyond the peak with a change from positive to negative sign at a radius of $\approx4.5$~kpc means the following for the S-shaped spiral galaxy M51a.

1) The corotation resonance position, related to the wave nature of the northern Spiral~2 is localised at a distance of $\approx4.5\pm0.15$ kpc ($118\arcsec$) from the galaxy centre,
which is compatible with the result of fitting the spiral pattern of M51a proposed by \citet{Tully1974} using density wave theory at the ''standard'' Toomre's stability criterion $Q=1.0$, $R_C=4.7$ kpc ($123\arcsec$), 
with the result of \citet{Zimmer2004} who applied the Tremaine-Weinberg method (TW-method) to the CO map of M51 and obtained an estimate of $R_C=4.2$ kpc ($110\arcsec$), with corotation resonances determined from reversals in the radial streaming motion caused by the spiral density wave \citep{Vogel1993,Font2024} and estimations $R_C=4.3-4.7$ kpc ($113-123\arcsec$) obtained in \citet{Abdeen2020}, using the method of overlaying images of galaxies observed at different wavelengths (see Table~\ref{table:summary} and Figure~\ref{figure:consistency}).

2) The positive sign of the azimuthal offset in the inner segment of the Spiral~2 means that extremely young OB-stars surrounded by H\,{\sc ii} regions are on average closer to the inner, concave edge of northern Arm~2 in SC--H{\sc ii}R pairs than relatively older star clusters without ionized gas, while the negative sign observed in the outer segment of Spiral~2 indicates that a subgroup of extremely young stars in SC--H{\sc ii}R pairs are on average closer to the outer, convex edge of northern Spiral~2.
This means that the matter in the disc inside the corotation radius ($R<4.5$ kpc) rotates faster than the northern Spiral~2 and, by catching up with it, forms a spiral shock on the inner, concave side of Arm~2, while the disc outside the corotation radius ($R>4.5$ kpc) rotates slower than Spiral~2 and forms a spiral shock on the outer, convex side of Arm~2.
Such a mutual arrangement of spiral pattern and stationary spiral shock indicates an anti-clockwise rotation of the matter in the disc.
The relation between the direction of galaxy rotation on the sky plane and the mutual arrangement of the spiral pattern and the stationary spiral shock is nicely illustrated in fig.~1 of \citet{Puerari1997}. 

\begin{figure}
\vspace{1.0mm}
\centerline{\includegraphics[width=0.48\textwidth]{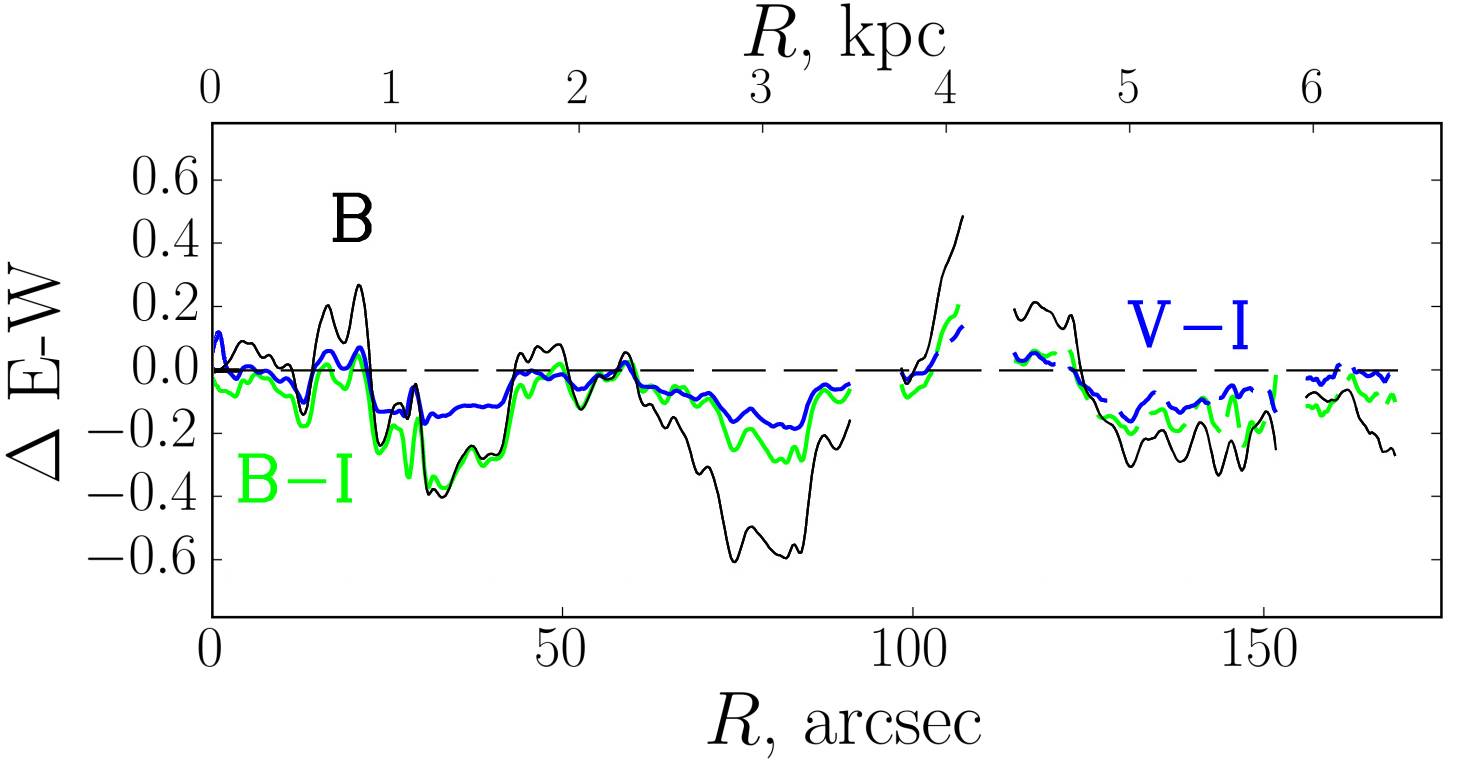}}
\caption{Photometric profiles of the difference in mean surface brightness in the $B$-band and the colour indices $B-I$, $V-I$ between the eastern (E) and western (W) parts along the minor axis in M51 (PA$=80\degr$). Averaging performed in a strip width of 1 arcmin.}
\label{figure:photometric_sections}
\end{figure}

The mutual position of the nearest (i.e., closest to the observer) eastern end of the minor axis (see Fig.~\ref{figure:photometric_sections}) and the observed location of the approaching half galaxy in the north of the sky plane \citep{Shetty2007,Colombo2014} provides independent confirmation of the counterclockwise rotation of the disc.
Figure~\ref{figure:photometric_sections} plots the photometric profiles of the difference in mean surface brightness in the $B$-band (in magnitudes per arcsec  squared) and the colour indices $B-I$ and $V-I$ between the opposite sides from the centre along the minor axis in M51, where 
\begin{eqnarray}
\Delta B_{\rm E-W}(R)=\mu(B)_{\rm E}(R)-\mu(B)_{\rm W}(R), \nonumber \\
\Delta(B-I)_{\rm E-W}(R)=(B-I)_{\rm E}(R)-(B-I)_{\rm W}(R), \nonumber \\
\Delta(V-I)_{\rm E-W}(R)=(V-I)_{\rm E}(R)-(V-I)_{\rm W}(R). \nonumber
\end{eqnarray}

The negative $\Delta B$, $\Delta(B-I)$, $\Delta(V-I)$ values in the figure indicate that the eastern side of the galaxy is brighter and bluer than the western side. Thus, the eastern side of M51a is the closest one to the observer. When viewed along the minor axis from the observer's nearest eastern edge towards the distant western end of the minor axis (galaxy viewed from above), the northern, approaching half (blueshifted side) of the disc is to the left of the minor axis, and the southern, receding half (redshifted side) is located on the right side of the minor axis \citep{Shetty2007,Colombo2014}. 
If the left side of the galaxy appears blueshifted and the right side is redshifted, the galaxy is rotating counterclockwise when viewed from above.
This is an independent evidence of the counterclockwise rotation of M51 on the sky plane derived from the radial variation of the median azimuthal offset in SC--H{\sc ii}R pairs in the northern Arm~2.

3) The S-shaped and counter-clockwise rotating northern Arm~2 points to a trailing spiral.

The radial profile of the azimuthal offset in the southern Arm~1 is also traceable starting from the region occupied by the end of the bar and shows a corotation resonance, $R_C=0.90\pm0.15$ kpc ($24\pm4\arcsec$), associated with the nuclear bar and which is confirmed by the aforementioned results of \citet{Querejeta2016,Zhang2012,Colombo2013,Meidt2008,Font2024}.
The further decrease of the mean azimuthal offset up to the negative peak at $R\approx2$ kpc ($\approx52\arcsec$) and its rise beyond the negative peak with a change from negative to positive sign means that a corotation resonance, related to the wave nature of the southern Spiral~1 is localised at a distance of $3.7\pm0.15$ kpc ($97\pm4\arcsec$) from the galaxy centre.

To reconcile the counterclockwise rotation of the disc with the radial profile of the mean azimuthal offset in the southern Arm~1, we must assume that in the inner disc, within the corotation resonance $R<4$ kpc, the counterclockwise rotation of Spiral~1 is faster than the rotation of matter in the disc.
The change of sign, from negative to positive, of the mean azimuthal offset at the passage through the corotation radius at $R=3.7\pm0.15$ kpc in the southern Spiral~1 suggests that outside this corotation circle the pattern speed is less than the anti-clockwise rotation velocity of matter in the disc, which, catching up with the southern Arm~1, forms a shock front from its inner, concave side.
Note that \citet{Meidt2008} investigating the evidence for radial variation in spiral pattern speed using the radial Tremaine-Weinberg (TWR) method found evidence for a decreasing of the pattern speed beyond the galactocentric distance $R>3.3$~kpc.

The different angular speeds of southern Spiral~1 and northern Spiral~2 indicate the presence of a complex multimode spiral system, which is predicted by studies of the radial amplitude profile of the spiral pattern in the $B$ and $I$ bands \citep{Elmegreen1989}, as well as by computer-enhanced images of the galaxy M51 \citep{Elmegreen1992}, in 
which the authors revealed evidence for the three arm spiral between $50\arcsec$ and $100\arcsec$ (1.9 and 3.8 kpc). 
\citet{Rix1993}, by Fourier analysis of the surface brightness in different photometric bands, identified a $m=3$ mode coexisting in the inner part of M51 with a dominant two-armed spiral.
Despite the strongly expressed two-arm spiral pattern in the M51, the three-arm pattern in it is very weak and does not clearly show itself in images of this galaxy.
According to \citet{Rix1993}, the features with $m=3$ are most likely to be present only in the dust and the gas, but not in the stars.
\citet{Elmegreen1992} suggested that this galaxy was only recently affected by tidal forces (which explains its two-armed structure), and it is possible that, like other galaxies, it will eventually form the strong three-armed structure we observe in other galaxies.
More recently, the presence of a multi-modal spiral system in M51 has been confirmed in \citet{Font2024} which found, by studying the kinematics of radial streaming gas motions in spiral arms, the existence of several modes with their own pattern speed.
Note that in analysing the velocity field of the molecular gas with high angular resolution in the framework of the PdBI Arcsecond Whirlpool Survey (PAWS\footnote{http://www.mpia.de/PAWS}), \citet{Colombo2013} obtained kinematic evidence for the existence of a $m=3$ mode of the gravitational density wave spatially coinciding with the inner part of the dominant two-arm spiral.
\citet{Meidt2008} considered the possibility that the weak $m=3$ structure identified photometrically in \citet{Rix1993}
in the inner part of the disk down to $R\sim2$ kpc ($\sim50\arcsec$), impacts the conclusions they obtained using the Tremain-Weinberg radial method (TWR).
Using CO(1-0) as a kinematic tracer, they found evidence of radial decreasing in the spiral pattern speed in M51a from an inner $\Omega_{p,1} = 90\pm25$ km/s/kpc out to $R=1.75\pm0.25$ kpc ($46\pm7\arcsec$) to an outer $\Omega_{p,2} =50\pm10$ km/s/kpc out to $R=3.4$ kpc ($89\arcsec$).
In addition, pattern speed, extending beyond 3.4 kpc out to 4.4 kpc is lower than the global $\sim38$ km/s/kpc found in \citet{Zimmer2004}. \citet{Meidt2008} proposed that their TWR pattern speed estimates, reflects the combination of a $m=3$ mode with a dominant two-armed spiral.

The cardinally opposite radial profiles of the mean azimuthal offset in the northern and southern arms could be explained 
within the framework of the hypothesis of the existence of two parallel counter-rotating discs in M51.
Within this hypothesis, the radial profile of the mean azimuthal offset in the northern arm~2 is the result of the interaction of matter 
in the disc, which rotates counterclockwise, with the northern spiral pattern, which also rotates counterclockwise, 
with  northern spiral~2 being the trailing spiral \citep[see Fig. ~1b in the][]{Puerari1997}.
While the interaction of matter in the second, oppositely clockwise rotating disc with a southern spiral pattern,  which is also solidly rotating clockwise, 
results to the observed radial profile  of the mean azimuthal offset in the southern arm~1,  with spiral~1 being the leading spiral \citep[see Fig.~1a in the][]{Puerari1997}.
It is only an assumption that needs further theoretical studies and observational confirmation.
So far, the hypothesis of opposite rotation of the spiral arms in galaxy M51 has been considered in several papers, mainly in the context of the galaxy's interaction with its companion galaxy NGC 5195.

\citet{Theis2003} conducted numerical simulations of NGC 5194's interaction with its companion galaxy NGC 5195, using an algorithm that mimics natural evolution under the assumption that the ''population'' improves its adaptation to the constraints of the ''environment'' during evolution.
They constrained the interaction by using the H\,{\sc i} intensity and velocity maps and recovered the counter-rotating H\,{\sc i}~arm, showing that tidal forces from NGC 5195 can induce irregular rotation patterns, confirming the possibility of opposing motions in different parts of NGC 5194's spiral structure.
Previously, \citet{Howard1990} presented an analytical analysis of numerical simulations of excitation mechanisms for M51 spiral arms, which suggests that, while the inner spiral segments have a wave nature triggered by tidal forces, the outer spiral segments manifest themselves as material arms, where gas clouds remain in the arm as the arm winds up. This leads to complex dynamics of spiral arms, including the possibility of counter-rotating structure.

\section{Conclusion}
\label{sect:conclusion}

Analysis of radial variations of the mean azimuthal offset between young star clusters and nearby H\,{\sc ii} regions in the northern and southern arms of M51 adds to the picture of the complex structure of this grand-design spiral galaxy, which has been revealed through numerous studies conducted by different authors using different methods over the last fifty years.
The totality of the results obtained points to the wave nature of the spiral structure of the M51a, which is the interference of the second mode ($m=2$) of the spiral density wave with the $m=3$ mode.

The cardinally opposite radial profiles of mean azimuthal offset in the northern and southern arms obtained in our study complemented this overall picture, 
showing that while the radial profile of mean azimuthal offset in the northern Arm~2 is consistent with the predictions of stationary density wave theory for trailing spirals, 
the explanation of the radial profile in the southern Arm~1 requires a pattern velocity within the corotation circle higher than the rotational velocity of matter in the disc and a lower pattern speed beyond the corotation resonance.
The opposite behaviour of the properties of the northern and southern arms is evident not only in the radial profiles of azimuthal offsets between tracers of different ages, as seen in Fig.~\ref{figure:offset_M51} of the current paper and fig.~7 in \citet{Egusa2017}, but also in the distribution of residual velocities of molecular clouds in the spiral arms.
\citet{Braine2020} investigated the rotation directions of molecular clouds in the inner region of M51 within the corotation circle, where the rotation curve grows, causing prograde rotation in gravitationally collapsing clouds in the inter-arm space, while a non-asymmetric potential, due to a spiral density wave, can cause retrograde rotation in clouds formed in the arms.
The mutual location of prograde and retrograde clouds relative to the concave and convex sides of the arm is determined by the relative velocity of the spiral pattern and matter in the disc. 
In case the two-arm spiral pattern rotates as a single solid body, the mutual location of the pro- and retrograde clouds relative to the concave and convex sides in the both arms will be symmetrical. 
If the pattern speed in one of the spirals is less than the rotational velocity of the disc, and the pattern speed in the other spiral is higher than the rotational velocity of the disc, the mutual location of the pro- and retrograde clouds relative to the concave and convex sides in the arms will be anti-symmetric, as shown in fig.~7 of \citet{Braine2020}.

Despite the different radial profiles of the mean azimuthal offset in the two arms, they are common by similar locations of corotation resonances and the magnitude of the maximum amplitude of the average azimuthal offset, confirmed by independent studies using different methods (see Table~\ref{table:summary} and Figure~\ref{figure:consistency}).
The negative peak of the amplitude of the azimuthal offset in the southern Arm~1 at a distance $R=2.10\pm0.15$ kpc ($54\pm4\arcsec$) corresponds to the position of the nuclear bar outer Lindblad resonance at $R=50\pm5\arcsec$ and the ultra-harmonic spiral resonance at $R=55\pm5\arcsec$, obtained in \citet{Querejeta2016} from the analysis of radial profiles of azimuthally averaged gravitational torques (see dark blue arrows at central and bottom panels of Figure~\ref{figure:offset_M51}).
The positive peak of the amplitude of the azimuthal offset in the northern Arm~2 at a distance $R=2.44\pm0.15$ kpc ($64\pm4\arcsec$) is in the region of a weak kink in the spiral pattern and the position of the density wave resonance indicator at $R=2.49\pm0.15$ kpc ($65.2\pm3.7\arcsec$) determined in \citet{Font2024} from reversals in the streaming motion caused by the density wave.

Comparing morphology of the arms with radial profiles of mean azimuthal offset shows qualitative agreement between the positions of kinks on radial profiles of azimuthal offset with the location of kink regions visible in the northern and southern arms.
The main kinks measured at somewhat different radial and azimuthal positions in each arm qualitatively coincide with the locations of kinks on the radial profiles of the amplitude of azimuthal offset in northern and southern arms at galactocentric distances $R\approx5.5$ kpc ($144\arcsec$) and $R\approx7$ kpc ($185\arcsec$), respectively and associated with the density wave resonances at $R=160.2\pm1.7\arcsec$ and $R=188.7\pm5.6\arcsec$ determined from reversals in the streaming motions \citep{Font2024}. Note that the positions of these kinks are also consistent, within the error bars, with the locations of the ultraharmonic and outer Lindblad resonances of the spiral pattern (see the orange and pink arrows in Figure~\ref{figure:offset_M51}).
These kinks in the radial profiles of the amplitude of the azimuthal offset in the northern and southern arms are probably related to the abrupt change, starting from $R>5.5$ kpc ($>145\arcsec$) outwards, of the pitch angle of the spiral pattern \citep{Egusa2017}, which in turn relates to the galaxy disc warping revealed by \citep[see][and references therein]{Oikawa2014}, from the observed anomalous fast drop (faster than Kepler's law) of the disc rotation rate at distances $R>6$ kpc.

Two groups of corotation estimates can be distinguished in the Table~\ref{table:summary} and Figure~\ref{figure:consistency}, which compares the corotation radius estimates obtained in the current paper with those obtained by other methods.

The first group includes estimates of the corotation radius lying in the area of influence of the bar, which radius is $R_{bar}=20\arcsec$ \citep{Comeron2010}.
Here we note that the $m=3$ mode ($R_{C, m=3}\approx30\arcsec$) detected in \citet{Colombo2013} from molecular gas kinematics, can provide an explanation for the dynamical coupling of the bar to the spiral structure as well as the position of the inner star-forming ring found in \citet{Comeron2014} 
around $R\approx20\arcsec$.

The second group of corotation estimates relates to southern and northern spiral, in which there is evidently a small radial offset between the corotation resonances. 
In the southern Arm~1, the spiral corotation takes place at a distance $R=3.70\pm0.15$ kpc ($97\pm4\arcsec$), while in the northern Arm~2 the spiral corotation is somewhat further from the centre at $R=4.50\pm0.15$ kpc ($118\pm4\arcsec$).
Our estimates of corotation radii in the southern and northern arms are consistent with the results of a number of independent studies carried out by different methods (see Table~\ref{table:summary}).

Finally, we note that the overlap of the internal corotational radius with the corotational resonance of the third mode ($m=3$) of the density wave, along with the strongly pronounced two-arm pattern, indicates the interference of the second and third modes of the density wave and the multi-modal nature of the spiral structure of M51 noted in \citet{Elmegreen1989, Marchuk2024}.
The coincidence of the positions of the ultraharmonic, Lindblad, and corotation resonances (also known as ''resonance coupling'') is an indication of the complex multi-modal entity of the spiral structure of M51, predicted by several theoretical models \citep{Sygnet1988, Baba2013, Rautiainen1999, Roskar2012, Minchev2012}, and revealed in some galaxies \citep{Sakhibov1987, Elmegreen1989, Meidt2008, Buta2009, Kostiuk2024, Font2014, Sakhibov2021, Marchuk2024AA}.

The opposite radial dependencies of the amplitude of the azimuthal offset in the northern and southern arms of M51, directly or indirectly confirmed by the results of several independent studies, should be encouraging for their authors and motivate them to further detailed observational studies and theoretical modelling of the properties and morphology of the M51 galaxy.

\section*{ACKNOWLEDGMENTS}

The authors thank A.~A.~Marchuk (Central (Pulkovo) Astronomical Observatory of the Russian Academy of Sciences) for his helpful and constructive comments. This study used open data from the NASA/IPAC Extragalactic Database 
(http://ned.ipac.caltech.edu), the PdBI Arcsecond Whirlpool Survey (PAWS; http://www.mpia.de/PAWS), and SExtractor software (http://sextractor.sourceforge.net).

\section*{FUNDING}

The study was conducted under the state assignment of Lomonosov Moscow State University. VSK thanks the Fund for the Development of Theoretical Physics and Mathematics ''BAZIS'' (project no.~23-2-2-6-1) for support.

\section*{CONFLICT OF INTEREST}

The authors of this work declare that they have no conflicts of interest.



\begin{thebibliography}{}

\bibitem [Abdeen et al.(2022)]{Abdeen2022}
Abdeen, S., Davis, B. L., Eufrasio, R., et al. 2022, Monthly Notices Royal Astron. Soc., 512, 366
\bibitem [Abdeen et al.(2020)]{Abdeen2020}
Abdeen, S., Kennefick, D., Kennefick, J., et al. 2020, Monthly Notices Royal Astron. Soc., 496, 1610
\bibitem [Baba et al.(2013)]{Baba2013}
Baba, J., Saitoh, T. R., \& Wada, K. 2013, Astrophys. J., 763, 46
\bibitem [Beckman \& Cepa(1990)]{Beckman1990}
Beckman, J. E. \& Cepa, J. 1990, Astron. and Astrophys., 229, 37
\bibitem [Berdnikov(1987)]{Berdnikov1987}
Berdnikov, L. N. 1987, Soviet Astronomy Letters, 13, 45
\bibitem [Bertin et al.(1989a)]{Bertin1989a}
Bertin, G., Lin, C. C., Lowe, S. A., \& Thurstans, R. P. 1989a, Astrophys. J., 338, 78
\bibitem [Bertin et al.(1989b)]{Bertin1989b}
Bertin, G., Lin, C. C., Lowe, S. A., \& Thurstans, R. P. 1989b, Astrophys. J., 338, 104
\bibitem [Braine et al.(2020)]{Braine2020}
Braine, J., Hughes, A., Rosolowsky, E., et al. 2020, Astron. and Astrophys., 633, A17
\bibitem [Buta \& Zhang(2009)]{Buta2009}
Buta, R. J. \& Zhang, X. 2009, Astrophys. J. Suppl., 182, 559
\bibitem [Colombo(2013)]{Colombo2013}
Colombo, D. 2013, PhD thesis, Ruprecht-Karls University of Heidelberg, Germany
\bibitem [Colombo et al.(2014)]{Colombo2014}
Colombo, D., Meidt, S. E., Schinnerer, E., et al. 2014, Astrophys. J., 784, 4
\bibitem [Comer{\'o}n et al.(2010)]{Comeron2010}
Comer{\'o}n, S., Knapen, J. H., Beckman, J. E., et al. 2010, Monthly Notices Royal Astron. Soc., 402, 2462
\bibitem [Comer{\'o}n et al.(2014)]{Comeron2014}
Comer{\'o}n, S., Salo, H., Laurikainen, E., et al. 2014, Astron. and Astrophys., 562, A121
\bibitem [Dixon(1971)]{Dixon1971}
Dixon, M. E. 1971, Astrophys. J. , 164, 411
\bibitem [Dobbs et al.(2010)]{Dobbs2010}
Dobbs, C. L., Theis, C., Pringle, J. E., \& Bate, M. R. 2010, Monthly Notices Royal Astron. Soc., 403, 625
\bibitem [Efremov(1985)]{Efremov1985}
Efremov, Y. N. 1985, Soviet Astronomy Letters, 11, 69
\bibitem [Egusa et al.(2009)]{Egusa2009}
Egusa, F., Kohno, K., Sofue, Y., Nakanishi, H., \& Komugi, S. 2009, Astrophys. J., 697, 1870
\bibitem [Egusa et al.(2017)]{Egusa2017}
Egusa, F., Mentuch Cooper, E., Koda, J., \& Baba, J. 2017, Monthly Notices Royal Astron. Soc., 465, 460
\bibitem [Egusa et al.(2004)]{Egusa2004}
Egusa, F., Sofue, Y., \& Nakanishi, H. 2004, Publ. Astron. Soc. Japan, 56, L45
\bibitem [Elmegreen(2011)]{Elmegreen2011}
Elmegreen, B. G. 2011, in EAS Publications Series, Vol. 51, EAS Publications Series, ed. C. Charbonnel \& T. Montmerle, 45–58
\bibitem [Elmegreen et al.(1992)]{Elmegreen1992}
Elmegreen, B. G., Elmegreen, D. M., \& Montenegro, L. 1992, Astrophys. J. Suppl., 79, 37
\bibitem [Elmegreen et al.(1989)]{Elmegreen1989}
Elmegreen, B. G., Elmegreen, D. M., \& Seiden, P. E. 1989, Astrophys. J., 343, 602
\bibitem [Elmegreen \& Lada(1977)]{Elmegreen1977}
Elmegreen, B. G. \& Lada, C. J. 1977, Astrophys. J., 214, 725
\bibitem [Elmegreen et al.(2011)]{Elmegreen2011b}
Elmegreen, D. M., Elmegreen, B. G., Yau, A., et al. 2011, Astrophys. J., 737, 32
\bibitem [Font et al.(2024)]{Font2024}
Font, J., Beckman, J. E., Epinat, B., Dobbs, C. L., \& Querejeta, M. 2024, Astrophys. J., 966, 110
\bibitem [Font et al.(2011)]{Font2011}
Font, J., Beckman, J. E., Epinat, B., et al. 2011, Astrophys. J., 741, L14
\bibitem [Font et al.(2014)]{Font2014}
Font, J., Beckman, J. E., Querejeta, M., et al. 2014, Astrophys. J. Suppl., 210, 2
\bibitem [Foyle et al.(2011)]{Foyle2011}
Foyle, K., Rix, H. W., Dobbs, C. L., Leroy, A. K., \& Walter, F. 2011, Astrophys. J., 735, 101
\bibitem [Fujimoto(1968)]{Fujimoto1968}
Fujimoto, M. 1968, Astrophys. J. , 152, 391
\bibitem [Gittins \& Clarke(2004)]{Gittins2004}
Gittins, D. M. \& Clarke, C. J. 2004, Monthly Notices Royal Astron. Soc., 349, 909
\bibitem [Gusev \& Shimanovskaya(2019)]{Gusev2019}
Gusev, A. S. \& Shimanovskaya, E. V. 2019, Monthly Notices Royal Astron. Soc., 488, 3045
\bibitem [Gusev et al.(2018)]{Gusev2018}
Gusev, A. S., Shimanovskaya, E. V., Shatsky, N. I., et al. 2018, Open Astronomy, 27, 98
\bibitem [Guti{\'e}rrez et al.(2011)]{Gutierrez2011}
Guti{\'e}rrez, L., Beckman, J. E., \& Buenrostro, V. 2011, Astron. J., 141, 113
\bibitem [Honig \& Reid(2015)]{Honig2015}
Honig, Z. N. \& Reid, M. J. 2015, Astrophys. J., 800, 53
\bibitem [Howard \& Byrd(1990)]{Howard1990}
Howard, S. \& Byrd, G. G. 1990, in NASA Conference Publication, Vol. 3098, NASA Conference Publication, ed. J. W. Sulentic, W. C. Keel, \& C. M. Telesco, 577–581
\bibitem [Kendall et al.(2015)]{Kendall2015}
Kendall, S., Clarke, C., \& Kennicutt, R. C. 2015, Monthly Notices Royal Astron. Soc., 446, 4155
\bibitem [Kendall et al.(2011)]{Kendall2011}
Kendall, S., Kennicutt, R. C., \& Clarke, C. 2011, Monthly Notices Royal Astron. Soc., 414, 538
\bibitem [Kennicutt et al.(2003)]{Kennicutt2003}
Kennicutt, Robert C., J., Armus, L., Bendo, G., et al. 2003, Publ. Astron. Soc. Pacific, 115, 928
\bibitem [Kostiuk et al.(2024)]{Kostiuk2024}
Kostiuk, V. S., Marchuk, A. A., \& Gusev, A. S. 2024, Research in Astronomy and Astrophysics, 24, 075007
\bibitem [Lin \& Shu(1964)]{Lin1964}
Lin, C. C. \& Shu, F. H. 1964, Astrophys. J., 140, 646
\bibitem [Lindblad(1963)]{Lindblad1963}
Lindblad, B. 1963, Stockholms Observatoriums Annaler, 5, 5
\bibitem [Louie et al.(2013)]{Louie2013}
Louie, M., Koda, J., \& Egusa, F. 2013, Astrophys. J., 763, 94
\bibitem [Marchuk(2024)]{Marchuk2024AA}
Marchuk, A. A. 2024, Astron. and Astrophys., 686, L14
\bibitem [Marchuk et al.(2024b)]{Marchuk2024}
Marchuk, A. A., Chugunov, I. V., Gontcharov, G. A., et al. 2024b, Monthly Notices Royal Astron. Soc., 528, 1276
\bibitem [Marchuk et al.(2024a)]{Marchuk2024b}
Marchuk, A. A., Mosenkov, A. V., Chugunov, I. V., et al. 2024a, Monthly Notices Royal Astron. Soc., 527, L66
\bibitem [Mart{\'\i}nez-Garc{\'\i}a et al.(2009)]{Martinez2009}
Mart{\'\i}nez-Garc{\'\i}a, E. E., Gonz{\'a}lez-L{\'o}pezlira, R. A., \& Bruzual-A, G. 2009, Astrophys. J., 694, 512
\bibitem [Mart{\'\i}nez-Garc{\'\i}a et al.(2023)]{Martinez2023}
Mart{\'\i}nez-Garc{\'\i}a, E. E., Gonz{\'a}lez-L{\'o}pezlira, R. A., \& Puerari, I. 2023, Monthly Notices Royal Astron. Soc., 524, 18
\bibitem [Meidt et al.(2008)]{Meidt2008}
Meidt, S. E., Rand, R. J., Merrifield, M. R., Shetty, R., \& Vogel, S. N. 2008, Astrophys. J., 688, 224
\bibitem [Miller et al.(2019)]{Miller2019}
Miller, R., Kennefick, D., Kennefick, J., et al. 2019, Astrophys. J., 874, 177
\bibitem [Minchev et al.(2012)]{Minchev2012}
Minchev, I., Famaey, B., Quillen, A. C., \& Dehnen, W. 2012, in European Physical Journal Web of Conferences, Vol. 19, European Physical Journal Web of Conferences, 07002
\bibitem [Miyamoto et al.(2014)]{Miyamoto2014}
Miyamoto, Y., Nakai, N., \& Kuno, N. 2014, Publ. Astron. Soc. Japan, 66, 36
\bibitem [Mueller \& Arnett(1976)]{Mueller1976}
Mueller, M. W. \& Arnett, W. D. 1976, Astrophys. J., 210, 670
\bibitem [Oikawa \& Sofue(2014)]{Oikawa2014}
Oikawa, S. \& Sofue, Y. 2014, Publ. Astron. Soc. Japan, 66, 77
\bibitem [Peterken et al.(2019)]{Peterken2019}
Peterken, T. G., Merrifield, M. R., Arag{\'o}n-Salamanca, A., et al. 2019, Nature Astronomy, 3, 178
\bibitem [Pettitt et al.(2017)]{Pettitt2017}
Pettitt, A. R., Tasker, E. J., Wadsley, J. W., Keller, B. W., \& Benincasa, S. M. 2017, Monthly Notices Royal Astron. Soc., 468, 4189
\bibitem [Pickel'ner(1971)]{Pickelner1971}
Pikel'ner, S. B. 1971, Sov. Astron., 14, 602
\bibitem [Pineda et al.(2020)]{Pineda2020}
Pineda, J. L., Stutzki, J., Buchbender, C., et al. 2020, Astrophys. J., 900, 132
\bibitem [Pour-Imani et al.(2016)]{Pour-Imani2016}
Pour-Imani, H., Kennefick, D., Kennefick, J., et al. 2016, Astrophys. J., 827, L2
\bibitem [Puerari \& Dottori(1997)]{Puerari1997}
Puerari, I. \& Dottori, H. 1997, Astrophys. J., 476, L73
\bibitem [Querejeta et al.(2016)]{Querejeta2016}
Querejeta, M., Meidt, S. E., Schinnerer, E., et al. 2016, Astron. and Astrophys., 588, A33
\bibitem [Rautiainen \& Salo(1999)]{Rautiainen1999}
Rautiainen, P. \& Salo, H. 1999, Astron. and Astrophys., 348, 737
\bibitem [Rix \& Rieke(1993)]{Rix1993}
Rix, H.-W. \& Rieke, M. J. 1993, Astrophys. J., 418, 123
\bibitem [Roberts(1969)]{Roberts1969}
Roberts, W. W. 1969, Astrophys. J., 158, 123
\bibitem [Ro{\v{s}}kar et al.(2012)]{Roskar2012}
Ro{\v{s}}kar, R., Debattista, V. P., Quinn, T. R., \& Wadsley, J. 2012, Monthly Notices Royal Astron. Soc., 426, 2089
\bibitem [Sakhibov et al.(2021)]{Sakhibov2021}
Sakhibov, F., Gusev, A. S., \& Hemmerich, C. 2021, Monthly Notices Royal Astron. Soc., 508, 912
\bibitem [Sakhibov \& Smirnov(1987)]{Sakhibov1987}
Sakhibov, F. K. \& Smirnov, M. A. 1987, Sov. Astron., 31, 132
\bibitem [Salo \& Laurikainen(2000)]{Salo2000}
Salo, H. \& Laurikainen, E. 2000, Monthly Notices Royal Astron. Soc., 319, 393
\bibitem [Savchenko et al.(2020)]{Savchenko2020}
Savchenko, S., Marchuk, A., Mosenkov, A., \& Grishunin, K. 2020, Monthly Notices Royal Astron. Soc., 493, 390
\bibitem [Scheepmaker et al.(2009)]{Scheepmaker2009}
Scheepmaker, R. A., Lamers, H. J. G. L. M., Anders, P., \& Larsen, S. S. 2009, Astron. and Astrophys., 494, 81
\bibitem [Seiden \& Gerola(1982)]{Seiden1982}
Seiden, P. E. \& Gerola, H. 1982, Fundamentals of Cosmic Physics, 7, 241
\bibitem [Shabani et al.(2018)]{Shabani2018}
Shabani, F., Grebel, E. K., Pasquali, A., et al. 2018, Monthly Notices Royal Astron. Soc., 478, 3590
\bibitem [Shetty et al.(2007)]{Shetty2007}
Shetty, R., Vogel, S. N., Ostriker, E. C., \& Teuben, P. J. 2007, Astrophys. J., 665, 1138
\bibitem [Shields et al.(2015)]{Shields2015}
Shields, D. W., Boe, B., Henderson, C. L., et al. 2015, in American Astronomical Society Meeting Abstracts, Vol. 225, American Astronomical Society Meeting Abstracts \#225, 250.09
\bibitem [Smirnov \& Sakhibov(1981)]{Smirnov1981}
Smirnov, M. A. \& Sakhibov, F. K. 1981, Akademiia Nauk Tadzhikskoi SSR Doklady, 24, 725
\bibitem [Sygnet et al.(1988)]{Sygnet1988}
Sygnet, J. F., Tagger, M., Athanassoula, E., \& Pellat, R. 1988, Monthly Notices Royal Astron. Soc., 232, 733
\bibitem [Tamburro et al.(2008)]{Tamburro2008}
Tamburro, D., Rix, H. W., Walter, F., et al. 2008, Astron. J., 136, 2872
\bibitem [Theis \& Spinneker(2003)]{Theis2003}
Theis, C. \& Spinneker, C. 2003, Astrophys. and Space Sci., 284, 495
\bibitem [Toomre(1974)]{Toomre1974}
Toomre, A. 1974, in IAU Symposium, Vol. 58, The Formation and Dynamics of Galaxies, ed. J. R. Shakeshaft, 347
\bibitem [Tully(1974)]{Tully1974}
Tully, R. B. 1974, Astrophys. J. Suppl., 27, 449
\bibitem [Vall{\'e}e(2020)]{Vallee2020}
Vall{\'e}e, J. P. 2020, New Astronomy, 76, 101337
\bibitem [Vogel et al.(1993)]{Vogel1993}
Vogel, S. N., Rand, R. J., Gruendl, R. A., \& Teuben, P. J. 1993, Publ. Astron. Soc. Pacific, 105, 666
\bibitem [Zhang \& Buta(2007)]{Zhang2007}
Zhang, X. \& Buta, R. J. 2007, Astron. J., 133, 2584
\bibitem [Zhang \& Buta(2012)]{Zhang2012}
Zhang, X. \& Buta, R. J. 2012, arXiv e-prints, arXiv:1203.5334
\bibitem [Zimmer et al.(2004)]{Zimmer2004}
Zimmer, P., Rand, R. J., \& McGraw, J. T. 2004, Astrophys. J., 607, 285

\end{thebibliography}


\end{document}